# Programmable Multifuctional Photonics ICs

Daniel Pérez, Ivana Gasullla and José Capmany:

*Photonics Research Labs*
*Universitat Politècnica de València*
Valencia, Spain
dperez@iteam.upv.es

*Abstract*— Here, we review present and future work in the next photonic IC generation aiming at integration of multi-functional software-defined systems for signal processing operations.

## I. INTRODUCTION

Since the first integration of photonic circuits, scientists and commercial partners have tried to increase the versatility of their components. This is commonly achieved by locally changing the effective index properties of the waveguides through the application of different electro-optical, thermo-optical, electro-mechanical or electro-magnetic effects. The controlled alteration of the phase and absorption of light has led to the design of integrated modulators, tunable filters, tunable splitters, optical switches and flexible beamforming networks, to cite a few examples. The notion of reconfigurable photonic components has evolved to more complex photonic integrated circuits (PICs), where the versatility is conferred to the whole system. By software programming each subsystem after fabrication, multiple functionalities are targeted with a single hardware platform, [1,2]. This new paradigm will radically enable cost-effective photonic-driven solutions to the market since the overhead costs of design, fabrication, packaging and testing stages can be shared between multiple applications. In addition, driving a reconfigurable hardware will create the expansion of a new field to be explored in terms of generic photonic physical architectures and software development opportunities.

Programmable Integrated Photonics can be divided into: Multipurpose devices and reconfigurable subsystems. While the latter can be reconfigured to increase and vary the performance of a certain application (e.g., filter tunability/ reconfigurability, programmable delays), the former spans on a wider range of applications. Here, we will provide an analysis and overview of the state of the art of programmable integrated photonics field.

## II. ARCHITECTURES

A reconfigurable photonic integrated circuit is typically built up by combining passive structures that allow light propagation, splitting and/or recombination together with structures that allow the modification of the phase of the optical signal based on an externally-driven actuator.

### A. Reconfigurable photonic integrated circuits

In practice, most of the reconfigurable photonic circuits proposed in the literature are either based on the cascade of finite (i.e., Mach-Zehnder interferometers) or infinite (i.e., ring cavity) impulse response cells or a combination of both. Some examples are illustrated in Fig. 1, which correspond to circuit configurations limited to the synthesis of 1-input/1-output or 2input/2-output circuits with fixed-period spectral response and delay line structures, [3]. As an example, most of the configurations can be designed to perform as optical filters that allow for bandwidth reconfiguration and notch tunability by moving their zeros and poles along the z-plane. In addition, they can be designed and configured to find applications in arbitrary dispersion compensation, optical equalization and multi-channel filter selection, [4]. Thus, these circuits provide a good compromise between performance, footprint and accumulated loss. However, once designed and fabricated, these might be limited to few applications and constrained to fixed free spectral ranges.

### B. Multifunctional programmable beamsplitting arrays

The great promise of versatile integrated optics lies in integrating the optimum number of components to perform the greatest number of functionalities with a common/shared hardware, [1]. In Fig. 2 (a), we can see an example of an integrated system that incorporates an array of electro/optic and opto/electronic converters, several optical sources among other high-performance blocks specifically designed to perform a single functionality (e.g., high-selectivity narrowband filtering and dispersive elements) and a reconfigurable optical core. These processing architectures allow multiple photonic processing functionalities as well as microwave / millimeter wave operations.

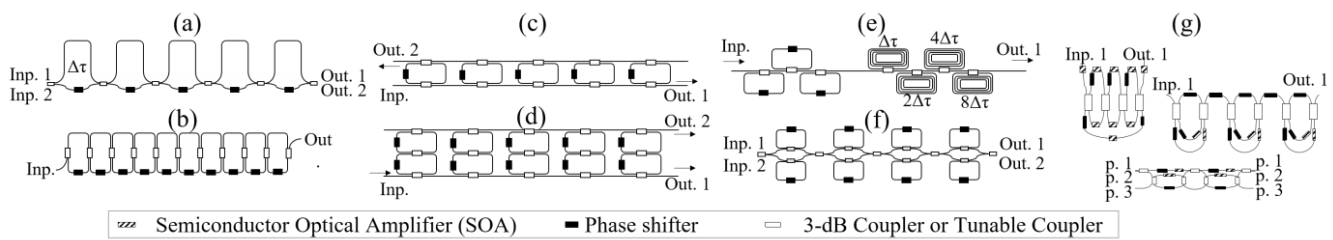

Figure 1. Schematic of common reconfigurable structures in photonic integrated circuits for optical processing: (a) Lattice filter of unbalanced Mach-Zehnder Interferometers (MZIs), [4] (b) Coupled Resonator Optical Waveguides (CROWs), [3], (c) side-coupled integrated spaced sequence of resonators (SCISSOR) with two bus waveguides, (d) Coupled CROW and SCISSORs, Single-bus SCISSORs, (e) tunable delay line, (f) Cavity-loaded cascade of MZIs, (g) Processing cells combining interferometric structures and SOAs.

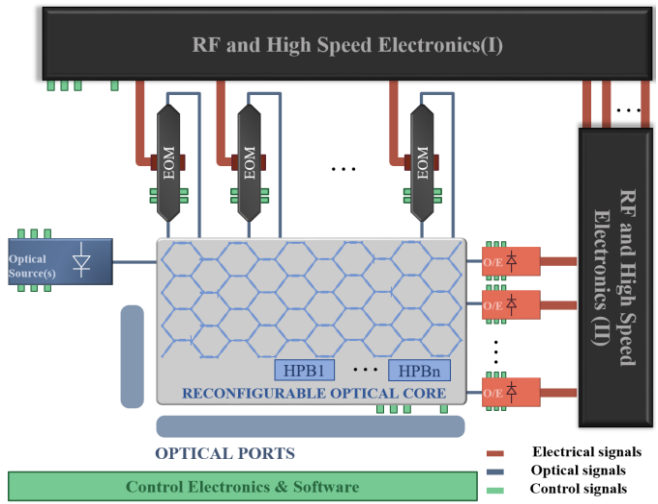

Figure 2. Schematic of a generic-purpose photonic processor architecture covering RF-photonic assisted operations and photonic operations.

A very versatile optical processing engine is required to implement the core of these architectures. The purpose of this element is two-fold. It needs, first, to provide a reconfigurable/dynamic interconnection scheme between all the subsystems. Secondly, it should provide the synthesis of the main optical signal processing operations. These tasks can be achieved by means of waveguide mesh arrangements [5-7]. These arrangements are massive interconnections of waveguide pairs arrays coupled by a 2x2 beamsplitting elements (mainly balanced Mach-Zehnder Interferometers with a phase modulator on each arm), describing vastly inter-coupled closed unity cells. Conventional PICs are then discretized into these beamsplitters or Tunable Basic Units (TBUs) to define the light path geometries and design parameters of the synthetized/programmed circuits. Once discretized, each TBU of the circuit is programmed either in cross-state, bar-state or as a tunable coupler with the desired splitting ratio. The common phase at each TBU can be set as well.

Depending on the optical node topology employed for the interconnection of the TBUs, the physical circuit will result in different waveguide mesh configurations. These topologies have been analyzed, highlighting the performance of the hexagonal waveguide interconnection topology as the most versatile and flexible, among other characteristics related to the integration itself, [7].

The first three experimental examples of this arrangement provide a powerful proof of concept of a new generation of programmable photonic integrated circuits. They are based on two square cells in silicon nitride [5], 30-TBUs describing a 7-cell hexagonal topology fabricated in silicon on insulator [6], and a recently fabricated hexagonal mesh based on 40 thermally-tuned TBUs currently under test.

Despite the simplicity of the layouts, even a 7-cell hexagonal structure is capable of implementing over 100 different circuits for optical filtering applications (basic MZI, finite impulse response transversal filters, basic tunable ring cavities and infinite impulse response filters, as well as compound structures such as CROWs and SCISSORs), true time delay lines and optical coherent interferometry, [6,8].

Once chosen a certain topology, the versatility of waveguide arrangements is proportional to the number TBUs. The example of Fig. 3 illustrates how each of the 81 TBUs of a waveguide mesh is configured to perform three PICs working in parallel simultaneously. Each one of the circuits consists of three optical ring resonators with three different cavity lengths. Additionally, the tunability of the shortest cavity is achieved by tuning the phase of the TBU labelled as C19. The response of each ring is computed based on a spectral analysis model, [9]. In this case, 0.2-dB insertion loss and 30-dB optical crosstalk are assumed for each TBU.

### III. WAVEGUIDE MESHES SCALABILITY: LIMITS AND CHALLENGES

If we compare the characteristics of classical reconfigurable PICs against the new generation based on massively coupled TBUs, we find that the versatility and the flexibility come at the cost of several limitations. It is clear that the ideal behaviour of the TBU leads to the perfect performance of the reconfigurable optical engine. However, in practice, several sources of degradation must be taken into account: imperfect splitting ratios, phase control, parasitic back-reflections, loss imbalances, fabrication errors (gradients through the circuit in thickness or temperature), and drift in time, [6]. Although these problems affect both circuit approaches, the ones based on waveguide mesh arrangements will be negatively fostered due to the inherent interconnection topology. We briefly discuss the most important ones below.

*Internal reflections*: The use of imperfect 3-dB couplers and/or possible fabrication errors that change the losses in the upper/lower arm of each TBU introduce optical crosstalk. Due to the cascade arrangement of TBUs, and the potential light recirculation of the mesh topologies, large optical crosstalk or a drift in each TBU configured state leads to signal leaking through the mesh. In [6], a TBU optical crosstalk below 30 dB was measured and maintained constant during the experimental demonstrations of programmed complex PICs, showing the robustness of the configured states. In addition, it can be showed that smart programming of the unused regions to guide the reflected and leakage signals as far as possible from the defined circuit or to defined drain optical ports, contribute to relax the specifications of the TBUs in term of optical crosstalk, [9].

*Accumulated losses*: The TBU insertion losses are one of the most limiting issues in optical mesh networks. Losses limit the maximum number of TBUs to define the programmed circuit and thus, the versatility and scalability of the overall mesh. In the same sense, a minimum TBU performance improvement leads to a high improvement in the overall mesh behaviour. State-of-the-art MZI-based TBUs insertion losses are below 0.25 dB in silicon photonics [10], which would allow a maximum number of 50 TBUs with a penalty of 10-dB loss in the programmed circuit.

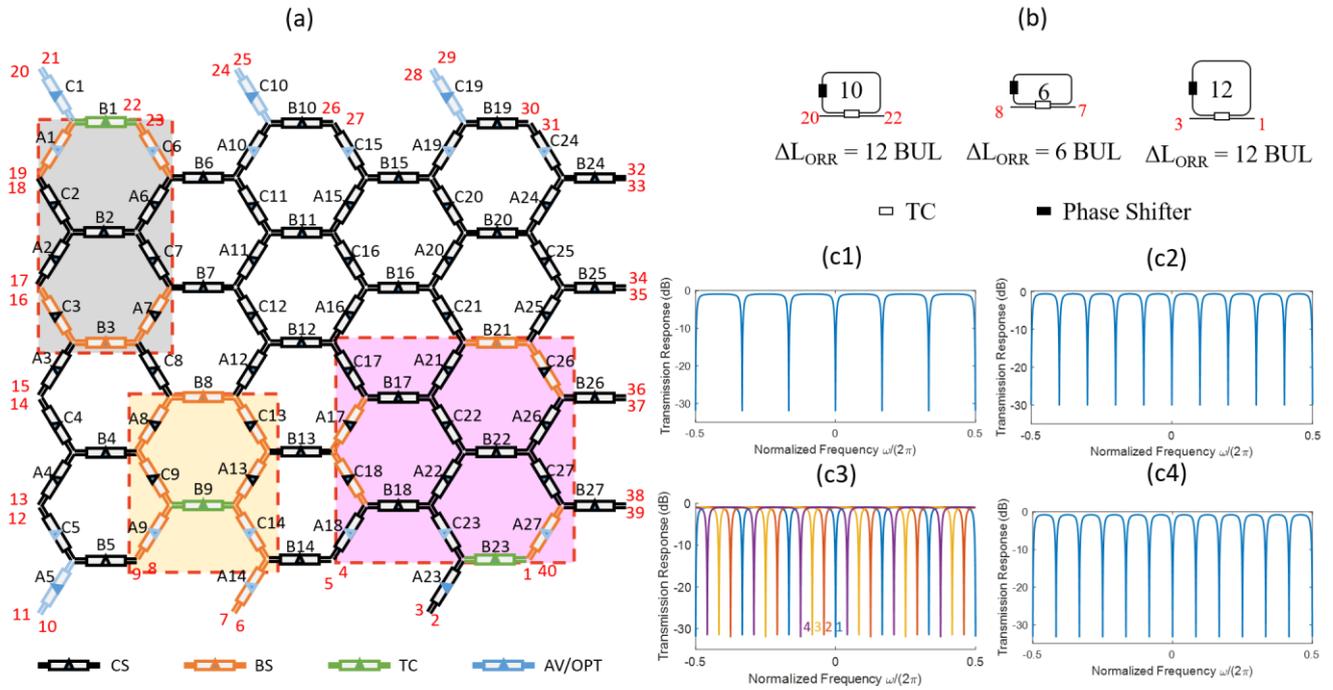

Figure 3. Spectral analysis of a waveguide mesh implementing three Optical Ring Resonator filters composed of cavity lengths equal to 6, 10, and 12 TBUs. (a) Mesh architecture and configuration for simultaneously implementing the three filters. (b) Equivalent circuit layouts with indication of the input and output ports in red ink (upper). (c1) Transmission response of the 6-BUL ORR, (c2) Transmission Response of the 10-BUL ORR and (c4) the 12-BUL ORR. (c3) Tunability response of 6-BUL ORR for $\phi_{c13}$ = 0, 0.5, 1 and 1.5, respectively

*Miniaturization trade-off: minimum delay and losses.* The structures susceptible of being programmed must be able to be discretized into a number of discrete TBUs. Ideally, a large count of TBUs with reduced length would enhance the resolution of the mesh allowing a better fitting of the circuits to be programed. A total BUL of 240 µm seems achievable with the state-of-the-art couplers and tuners. Assuming a typical SOI group index of 4.18, this is translated to maximum FSRs of around 150 and 50 GHz for the synthesis of MZIs and ORRs, respectively, in the hexagonal waveguide mesh topology, [7]. However, a reduction of the BUL implies that the signal must go through a greater number of TBUs to obtain a desired delay. If the 3-dB couplers limit the overall IL of the TBU, this miniaturization trade-off must be considered.

Finally, waveguide mesh arrangements face similar extended issues to the ones handled by the integrated optical switching matrices research field. Thermal stability, tuning-based crosstalk, power consumption, integration density and packaging involving a large count of electrical and optical i/o ports. However, the integration of less than 100 TBUs provides a considerable number of circuit topologies to be designed and limited, in principle, by accumulated IL in the largest circuit synthesis. The integration in silicon of more than 450 thermally tuned TBUs in optical switch networks has been recently demonstrated [11].

The current research in the field is focused on alternative tuning mechanisms to reduce the overall power consumption and footprints, TBU geometries and loss optimization, as well as the development of software to drive and control the reconfigurable subsystems. All in all, the arising of general-purpose programmable PICs is a promising candidate to enable cost-effective devices by a reduction of non-recurring engineering costs, shorter times to production and to market, as well as multifunctional and multitask operation.

ACKNOWLEDGMENT

The authors acknowledge financial support by the ERC ADG-2016 UMWP-Chip, the Generalitat Valenciana PROMETEO 2017/017 research excellency award, the COST Action CA16220 EUIMWP, and the Spanish MINECO Ramon y Cajal program RYC-2014-16247 for I. Gasulla's fellowship.